\documentclass[useAMS,usenatbib]{mn2e}
\usepackage{epsfig}
\title[The cool dust contents of powerful radio galaxies]{The dust masses of powerful radio galaxies: clues to the triggering of their activity}
\author[C. Tadhunter et al.]{C. Tadhunter$^{1}$\thanks{E-mail:
c.tadhunter@sheffield.ac.uk}, D. Dicken$^{2}$, R. Morganti$^{3,4}$, V. Konyves$^{2}$, N. Ysard$^{5}$, \newauthor N. Nesvadba$^{5}$, C. Ramos Almeida$^{6,7}$  \\
$^{1}$Department of Physics \& Astronomy, University of Sheffield, Sheffield
S3 7RH\\
$^{2}$Laboratoire AIM-Paris-Saclay, CEA/DSM/Irfu, Orme des Merisiers, Bat 709, 91191 Gif sur Yvette, France \\
$^{3}$ASTRON, P.O. Box 2, 7990 AA Dwingeloo, The Netherlands \\
$^{4}$Kapteyn Astronomical Institute, University of Groningen, P.O. Box 800,
9700 AV Groningen, The Netherlands \\
$^{5}$Institut d'Astrophysique Spatiale (IAS), UniversitŽ Paris-Sud, 91405, Orsay, France \\
$^{6}$Instituto de Astrof\'{i}sica de Canarias, C/ Via L\'{a}ctea, E38205 - La Laguna,Tenerife, Spain \\
$^{7}$Departamento de Astrof\'isica, Universidad de La Laguna, E-38205, La Laguna, Tenerife, Spain \\
}
\begin{document}

%\date{}

%\pagerange{\pageref{firstpage}--\pageref{lastpage}} \pubyear{2002}

\maketitle

\label{firstpage}

\begin{abstract} We use deep Herschel
Space Observatory observations of a 90\% complete sample of 32 intermediate-redshift 2Jy radio galaxies ($0.05 < z < 0.7$) to estimate the dust
masses of their host galaxies and thereby investigate the triggering mechanisms for their quasar-like AGN. The dust masses  
derived for the radio galaxies ($7.2\times10^5 < M_d < 2.6\times10^8$~M$_{\odot}$) are intermediate between those of quiescent elliptical galaxies on the one hand, and ultra luminous infrared galaxies (ULIRGs) on the other. Consistent with simple models for the co-evolution of  supermassive black holes and their host galaxies, these results suggest that most radio galaxies represent the late time re-triggering of AGN activity via mergers between the host giant elliptical
galaxies and companion galaxies with relatively low gas masses. However,
a minority of the radio galaxies in our sample ($\sim$20\%) have high, ULIRG-like dust masses, along with evidence for prodigious star formation activity. The latter objects are more likely to have been triggered in major, gas-rich mergers that represent a rapid growth phase for both their host galaxies and their supermassive black holes.

\end{abstract}


\section{Introduction}

%AGN are increasingly recognised as a key element in the evolution of galaxies
%of all types. Not only are AGN likely to be {\it triggered} as a consequence of gas %infall as galaxies build up their masses, but there is growing evidence that they %also directly affect the evolution of their host galaxies via the {\it feedback} %effects of their jets and winds \citep[e.g.][]{cattaneo09}. Clearly, if we are to %properly incorporate AGN into galaxy evolution models, it is crucial to understand %how and when AGN are triggered as their host galaxies evolve.

Radio-loud AGN are particularly important for understanding the link(s) between galaxy evolution and nuclear activity, because, in contrast to samples of radio-quiet AGN in the local Universe (e.g. Seyfert galaxies), they are invariably associated with early-type host galaxies, allowing relatively ``clean'' searches to be made for the signs of the triggering events. They also drive powerful relativistic jets that are capable of projecting the power of the AGN into the gaseous haloes of the host galaxies and clusters of galaxies, directly affecting the cooling of the hot IGM/ICM gas -- one of the most important forms of AGN-induced feedback \citep{best06,mcnamara07}.

Recently we have undertaken a major programme of optical \citep{ramos11,ramos12} and mid- to far-IR \citep{dicken08,dicken09,dicken12} observations of the 2Jy sample of southern radio galaxies \citep[$0.05 < z < 0.7$:][]{tadhunter98}, aimed at directly investigating the triggering mechanisms for radio-loud AGN. Our deep Gemini optical imaging observations show tidal features (tails, shells, fans, bridges) at relatively high surface brightness levels in a majority of objects ($>$85\%), consistent with the idea that radio-loud AGN represent a fleeting active phase in the evolution of the subset of giant elliptical galaxies that have recently undergone galaxy interactions and mergers \citep{ramos12}. However, the radio-loud AGN  phase does not correspond to a single stage of the merger sequence, nor to a single type of galaxy merger or interaction \citep{ramos11,ramos12}. Such apparent diversity is also consistent with the fact that only a minority ($\le$35\%) of our sample show evidence for recent star formation activity, based on the direct detection of young stellar populations at optical wavelengths \citep{tadhunter02,tadhunter11}, far-IR excesses from Spitzer photometry \citep{dicken09}, and  PAH features in mid-IR Spitzer spectroscopy \citep{dicken12}. 

%By comparing the space densities of radio-loud AGN with those of morphologically %disturbed elliptical galaxies in matched control samples, and taking into account %the relative timescales of the radio source activity (1 -- 100 Myr) and visibility %of the tidal features ($\sim$1 Gyr), 
%We have shown that 
Our previous work suggests that only a small fraction ($\le$20\%) of morphologically disturbed elliptical galaxies are capable of hosting a powerful radio-loud AGN \citep{ramos12}. Therefore, simply undergoing a galaxy interaction of {\it any type} is not by itself sufficient to trigger a powerful radio-loud AGN in a giant elliptical galaxy. One of the most important additional factors is likely to be gas content: it is probable that powerful, FRII-like radio activity is triggered by accretion of the cool/warm phases of the ISM, rather than the hot phase \citep[e.g.][]{hardcastle07}. Therefore we expect the ability to host a powerful radio-loud AGN to be related at some level to the total reservoir of cool ISM which, in the case of the elliptical galaxy hosts of radio-loud AGN, is likely to have had an external origin. 

Simple arguments allow us to predict the total gas reservoir required to trigger quasar-like activity
and sustain it over the quasar lifetime ($t_{qso}$ yr). If we define a quasar to have a bolometric luminosity of 
$L_{BOL} > 10^{38}$~W, a mass inflow rate of $\dot{M} > 0.2$~M$_{\odot}$~yr$^{-1}$ onto the central supermassive black hole is required to produce the quasar activity, assuming an efficiency of 10\%. Therefore over the lifetime of the quasar the supermassive black hole will accrete a mass $M_{ac} > 0.2 t_{qso}$~M$_{\odot}$, amounting to 
$M_{ac} > 2\times10^6$~M$_{\odot}$ for a typical quasar lifetime of $t_{qso} \sim 10^7$~yr \citep{martini04}. However, on the basis of the black hole mass vs host bulge mass correlations, we know that only a small fraction of the gas accreted by the bulge reaches the supermassive black hole: for every solar mass accreted by a supermassive black hole, $\sim$500 solar masses of stars must be formed in the bulge of the host galaxy \citep{marconi03}. In this case, the {\it total} gas reservoir that is expected to be associated with a particular quasar triggering event ($M_{tot}$) is much larger: $M_{tot} \sim 100 t_{qso}$~M$_{\odot}$ or $M_{tot} \sim 10^9$~M$_{\odot}$ for $t_{qso} \sim 10^7$~yr. Although this argument is over-simplified in the sense
that some of gas may already have been used up in star formation or ejected by feedback processes by the time the quasar is observed, it provides an order-of-magnitude estimate of the total gas reservoir required. 

\begin{figure} 
\epsfig{file=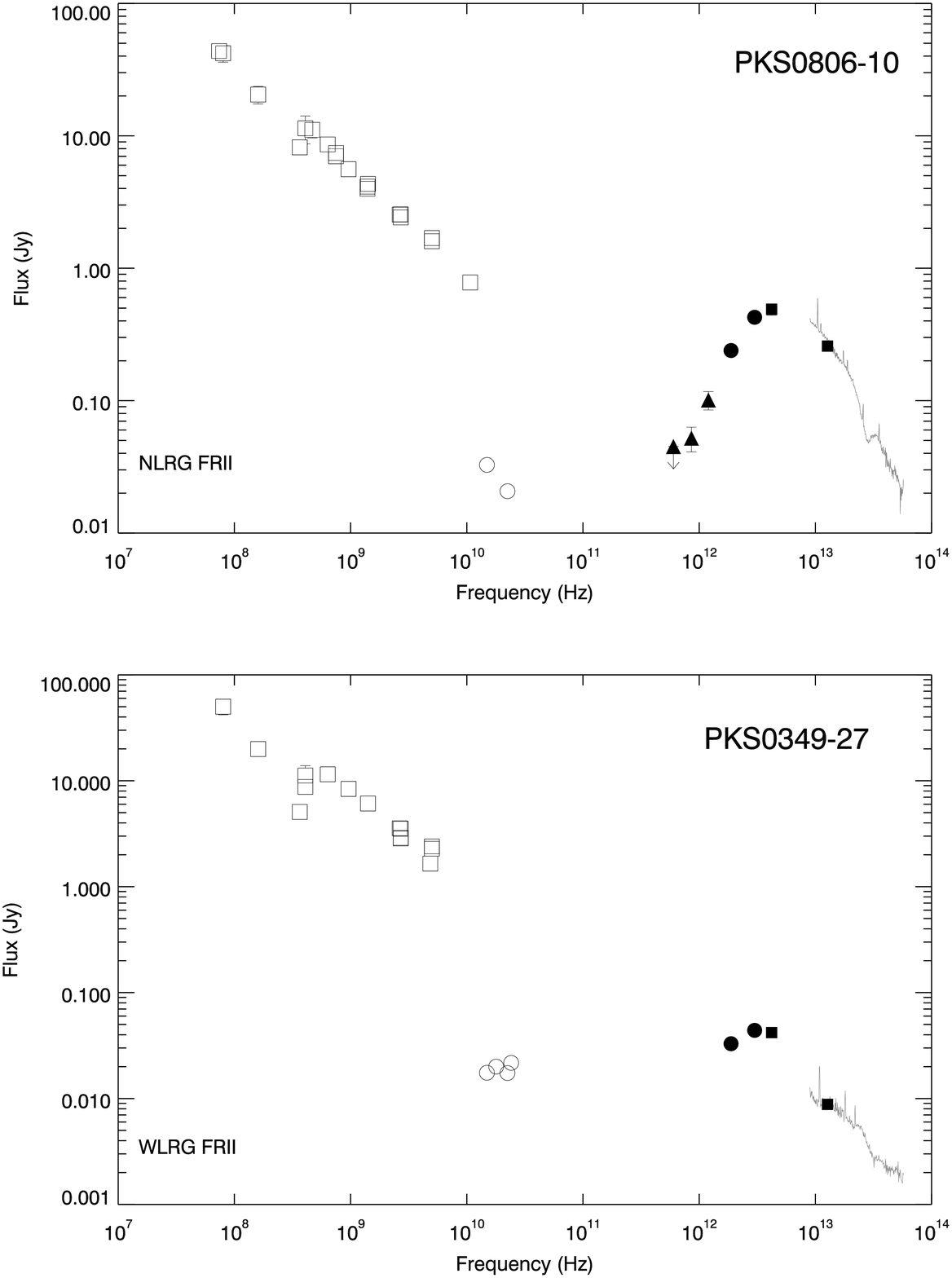,width=8cm}
\caption{Example radio to infrared SEDs for PKS0806-10 (top) and PKS0349-27 (bottom). The symbols are as follows. Open squares: total radio fluxes; open circles: core radio fluxes; filled triangles: Herschel/SPIRE fluxes; filled circles: Herschel/PACS fluxes; filled squares: Spitzer/MIPS fluxes. The solid lines represent the Spitzer/IRS spectra for the sources.}
\end{figure}

In this letter we use deep Herschel observations of a sample of nearby powerful radio galaxies with quasar-like nuclei to address the issue of whether the total cool ISM reservoirs are as massive as expected based on this simple model for the co-evolution of galaxy bulges and their central supermassive black holes. We also use the results to investigate the nature of the triggering merger events.
A cosmology with $H_0 = 71$~km s$^{-1}$ Mpc$^{-1}$, $\Omega_m=0.27$, and
$\Omega_{\lambda}=0.73$ is assumed throughout this paper.

\section{Sample, observations and data reduction}

The full sample for this project, as described in \citet{dicken09,dicken12}, comprises all 46 radio sources with redshifts $0.05 < z < 0.7$, radio fluxes
$S_{2.7GHz} > 2.0$~Jy, declinations $\delta < +10$~degrees, and radio spectral indices $\alpha < -0.7$ (assuming $F_{\nu} \propto \nu^{+\alpha}$). This southern 2Jy sample is now the best observed of all samples of powerful radio galaxies in terms of the depth and completeness of its multi-wavelength data\footnote{See http://2jy.extragalactic.info/The\_2Jy\_Sample.html}.

All 46 objects in the southern 2Jy sample were observed at 100 and 160$\mu$m with the Herschel PACS instrument as part of a programme of ``Must Do'' observations in the last days of Herschel; and the 16 objects with the brightest Spitzer 70$\mu$m fluxes were also observed with the Herschel SPIRE instrument at 250, 350 and 500$\mu$m in order to characterise the shapes of their longer-wavelength SEDs. The depths of the PACS observations allowed us to detect 100\% and
87\% of the sample at 100 and 160$\mu$m respectively. In addition, for the purposes of quantifying the degree of any non-thermal contamination of the Herschel photometry, we made sub-mm 870$\mu$m observations with
LABOCA/APEX of the 20 objects previous identified with Spitzer as having the possibility of non-thermal contamination \citep{dicken08}. Although our full sample comprises a mixture of weak-line radio galaxies (WLRG) and strong-line radio galaxies
(SLRG: $EW_{[OIII]} > 10$~\AA), in this paper we concentrate on the 35 objects in our sample with strong emission lines, indicating the presence of quasar-like AGN in addition to the powerful radio jets and lobes. Discussion of the full sample, including details of the observations, data reduction, far-IR photometry and SEDs, will be presented in \citet{dicken14}, while basic information about the SLRG sample
used in this paper is presented separately in the Appendix in Table~A1.

In Figure 1 (top) we show an example SED of one of the brighter sources in our sample for which we
have full flux information across the Herschel PACS and SPIRE bands, while in Figure 1 (bottom) we show
the SED of a more typical, fainter object that was only detected with PACS. 
%These SEDs incorporate
%Spitzer/IRS spectroscopy, Spitzer/MIPS MFIR photometry at 24 and 70$\mu$m, 
%LABOCA sub-mm photometry,
%and radio flux data.

Given that all the objects in our sample are radio-loud AGN, it is important to be cautious about the potential contamination of the far-IR continuum by non-thermal emission from the synchrotron-emitting radio jets, lobes and cores. As discussed in \citet{dicken08}, such contamination is most likely to be important in objects where the flat-spectrum cores and inner jets are unusually strong, or the steep-spectrum radio lobes are contained within the Herschel beam for the PACS and SPIRE observations. By using simple power-law extrapolations of our
high frequency (20~GHz) radio and LABOCA 870$\mu$m measurements, we find that non-thermal contamination is significant at far-IR wavelengths in 6/35 (17\%) of the objects in our SLRG sample  \citep[see][for details]{dicken14}. For three of these objects -- all compact steep spectrum sources -- it has proved possible to accurately subtract the non-thermal power-law, and thereby derive fluxes and luminosities for the thermal dust emission. However, for the remaining three sources (PKS0035-02, PKS1136-13 and PKS1306-09) the non-thermal
contamination makes a larger contribution at far-IR wavelengths, and we cannot confidently subtract this component because of uncertainties about the extrapolation of the SEDs. Therefore we exclude the
latter three objects from our analysis\footnote{Note that the upper limiting dust masses, derived for these objects by assuming that all their 160$\mu$m radiation is dust emission, fall well within the range measured for the other objects in our sample; excluding these objects does not affect any of the comparisons with the other samples in section 4.}, making our sample of 2Jy SLRG 90\% complete overall.

%In the latter cases, the far-IR continuum luminosities, and the dust masses derived from them, are 
%necessarily upper limits.

\section{Determing the cool dust masses}

The determination of dust masses from the far-IR data is highly sensitive to the
assumptions made in the analysis, for example the form of the long-wavelength dust
emissivity law, as characterised by the $\beta$ index for a modified black body fit
($F_{\nu} \propto \nu^{\beta} B(\nu,T)$), and the number of cool dust components of different temperature
assumed in the model (e.g. Clements, Dunne \& Eales 2010).

In order to simplify our analysis, and allow the direct comparisons with the dust masses of ULIRGs and
normal non-active elliptical galaxies presented in Section 4 and Figure 2, we assumed a single
temperature ($T_d$) for the cool dust in each object. Furthermore, we assumed that all objects in
our sample have the same
$\beta$ index. In this case the dust mass is given
by:
\begin{equation}
M_d = \frac{S^K_{160} D^2}{\kappa_{160}^m B(160,T_d) (1+z)}
\end{equation}
where $S^K_{160}$ is the K-corrected flux density at 160$\mu$m, $D$ is the luminosity distance,  $\kappa_{160}^m$ is the dust opacity per unit mass at a wavelength of 160$\mu$m, and $B_{\nu}(160,T_d)$ is the flux density of a black body of
temperature $T_d$ at 160$\mu$m. For the purposes of this analysis we assume 
$\kappa_{160} = 1.06$~m$^2$ kg$^{-1}$ \citep{draine03}.

We determined a typical $\beta$ index for the objects in our sample by comparing the 100/160 and 160/250
colours measured for the brighter objects observed with SPIRE, with the predicted colours for modified black bodies of different temperature and
beta index \citep[see][]{dicken14}. From this comparison we find that  $\beta = 1.2$ is most consistent with the observations of the far-IR bright 2Jy objects in our sample.

%\begin{figure} 
%\psfig{file=color_colour.ps,width=8cm}
%\caption{Far-IR colour-colour plot for objects in our sample that have
%good Spire 250$\mu$m data and for which which non-thermal contamination is not an issue (open symbols). The %filled symbols show the colours expected for a modified black body with $\beta = 1.2$ ands
%various temperatures and redshifts.
% }
%\end{figure}

We then determined the temperatures of the cool dust components for all the objects in our sample with both 100 and 160$\mu$m fluxes by comparing
the 100/160 flux ratios with those predicted by modified black body curves of different $T_d$ but
the same $\beta$ index ($\beta = 1.2$), with uncertainties in the temperatures estimated from the
uncertainties in the 100/160 flux ratios. The cool dust temperatures fall in the range $30 < T_d < 70$~K, with a median dust temperature $T_d = 41$~K. Finally, the dust masses were  calculated from the K-corrected
160$\mu$m fluxes and estimated temperatures using equation 1 above. Note that the uncertainties in the dust masses take into account 
the uncertainties in both the far-IR fluxes and the dust temperatures, but are inevitably dominated by the uncertainties in the temperatures.

Six of  the SLRG in our sample were not detected at 160$\mu$m. In such cases, we derived the dust masses using the K-corrected 100$\mu$m fluxes, assuming the median dust temperature derived from the
other 26 objects in the sample, and $\kappa_{100}^m = 2.71$~m$^2$ kg$^{-1}$ \citep{draine03} in equation 1. The distribution dust masses for these six objects is similar to that of the sample as a whole. Therefore the inclusion/exclusion
of these objects does not alter our conclusions.

Clearly, our dust mass results are sensitive to the assumed value of the $\beta$ index, which affects
both the temperatures derived from the 100/160 flux ratios and the K-correction of the far-IR fluxes.
However, we find that changing the $\beta$ index from 1.2 to 2.0 -- the value assumed in some studies 
of dust masses in elliptical galaxies, for example -- only increases the median dust mass by 80\%. Moreover, comparisons between the dust masses derived for various overlapping samples of low redshift ULIRGs \citep{klaas01,farrah03,clements10} and elliptical galaxies \citep{smith12,diserego13,martini13} suggest that the effect of changing the $\beta$ index, number of dust components, and general modelling approach is at most a
factor of $\sim$2 -- 3 in the dust mass. This will not significantly affect the main results of the comparisons with the samples of the quiescent elliptical galaxies and ULIRGs presented in the next section.

\section{Results}

The main result of this study is that the SLRG in our complete sample are moderately dust-rich. Figure 2 shows a 
histogram of the dust masses derived from our Herschel data. The dust masses cover the
range $7.2\times10^5 < M_d < 2.6\times10^8$~M$_{\odot}$, with a median dust mass of $8.5\times10^6$~M$_{\odot}$.
Assuming a typical gas-to-dust ratio of 140 \citep{draine07,parkin12}, these translate into gas masses in the
range $1.0\times10^8 < M_g < 3.7\times10^{10}$~M$_{\odot}$, with a median gas mass of $1.2\times10^9$~M$_{\odot}$.
We note that this median gas mass is remarkably close to that predicted on the basis of the simple argument presented in the introduction. Therefore, it appears that there is indeed sufficient cool ISM in these radio galaxies to sustain the quasar-like activity for the requisite timescales, and simultaneously grow
the stellar mass of the host galaxies, maintaining the black hole/host scaling relationships.

To put these results into context, Figure 2 also shows the distributions of dust masses derived for nearby ULIRGs and elliptical galaxies. The ULIRG dust masses were taken from the results of the variable $\beta$ index modified black body fits to the far-IR/sub-mm SEDs of the sample of 23 nearby ($z< 0.17$) ULIRGs presented by \citet{clements10}, while the elliptical galaxy dust masses were taken from the modified black body $\beta = 2$ results for the  non-active
%\footnote{By non-active we mean lacking a Seyfert or quasar AGN or a powerful FRII %radio source, but some objects with  LINER optical classification and low power FRI %radio activity are included.} 
ellipticals in \citet{smith12} and \citet{diserego13}, as well as from the template fitting results of Martini et al. (2013); the final sample of elliptical galaxies comprises 40 objects
with distances in the range $15 < D <66$~Mpc.
%\footnote{In the cases of objects 
%in common between the three papers on the elliptical galaxy dust masses, for each %object we took the highest dust mass estimate; this minimises any differences with %the radio galaxy dust masses, which tend to be higher.}. We note that a significant %proportion (53\%) of the objects in the nearby elliptical galaxy sample remain %undetected
%in sensitive far-IR observations, and therefore only upper limits on their dust %masses are available. However, we include these upper limits in the histograms of %Figure 2 as if they are
%detections. This makes our comparison with the dust masses of the 2Jy radio %galaxies conservative.
%Multiple modified black body fits --- including cooler and warmer dust components %--- tend to give higher dust masses. For example, in the case of %\citep{clements10}, changing from a one component modified black body fit (with %$\beta$ free) to a two component modified black body fit (with $\beta=2.0$ for both %components) increased the total dust mass by a median factor of $\sim$1.9 on %average. However, we would expect to see a similar increase in dust mass if we were 5to fit two component models to the 2Jy radio galaxy data. Therefore this does not %affect our conclusions.

The median dust masses of the ULIRG and elliptical galaxy samples plotted in Figure 2 are
$8\times10^7$~M$_{\odot}$ and $2\times10^5$~M$_{\odot}$ respectively.
At the high dust mass end of the distributions there is clearly an overlap between the dust massses of the radio galaxies and those of the ULIRGs; the lower end of the radio galaxy dust mass distribution
also overlaps with the higher end of the elliptical dust mass distribution. However, on average, the radio galaxies have dust masses that are more than an order of magnitude {\it higher} than those of typical elliptical galaxies, but an order of magnitude {\it lower} than those of typical ULIRGs. 

We emphasise that our comparison between the 2Jy radio galaxies and the elliptical galaxies is conservative, because (a) we have included upper limiting dust masses for the elliptical galaxies ($\sim$53\% of cases) as if they were detections; (b) for objects in common between the samples we have favoured the Smith et al. (2012) single modified black body fits --- which give generally larger dust masses --- over the Martini et al. (2013) template fitting results; and (c) the elliptical galaxy dust mass estimates are dominated by the results of the Smith et al. (2013) and di Serego Alighieri et al. (2013) $\beta=2.0$ single modified black body fits, which tend to give higher dust masses than the $\beta=1.2$ fits used for the radio galaxies. 

In terms of the comparison with the ULIRGs, the single modified black body fits of \citet{clements10} with variable beta index (as shown in Figure 2) are the most comparable with the $\beta=1.2$ results presented for the radio galaxies. However, even if we compare the $\beta=2.0$ results for the 2Jy radio galaxies with the $\beta=2.0$ results of \citet{klaas01} --- which give the lowest dust masses of all the low-z ULIRG studies --- the radio galaxy dust masses remain a factor $\sim$4 lower on average.

We further note that the total cool ISM mass implied by the median radio galaxy dust mass ($M_g = 1.2\times10^9$~M$_{\odot}$) is significantly lower than that of the Milky Way \citep[$4.8\times10^9$~M$_{\odot}$:][]{draine11}, and about a factor of two higher than that of the Large Magellanic Cloud (LMC) \citep[$5\times10^8$~M$_{\odot}$:][]{westerlund90}.

\begin{figure}
\epsfig{file=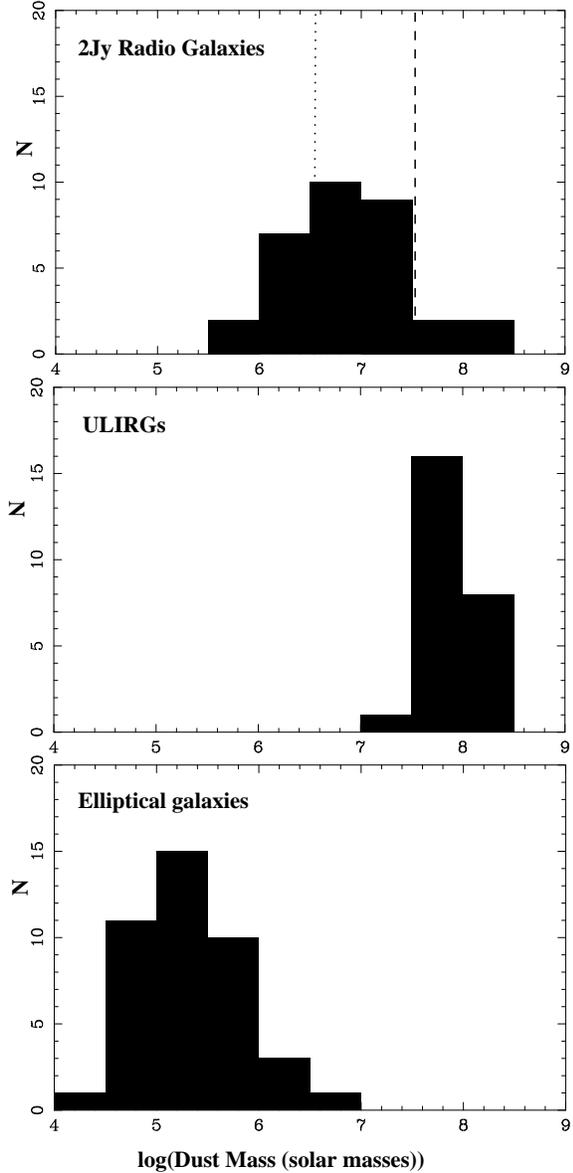,width=7.5cm}
\caption{Dust mass distributions for 2Jy radio galaxies (top), ULIRGs (middle), and elliptical
galaxies (bottom); see text for details of the sample selection. The vertical dashed and dotted lines show the equivalent dust masses of the Milky way and LMC respectively (both calculated from total hydrogen gas masses assuming $M_g/M_d = 140$).}
\end{figure}

The other interesting feature of the radio galaxy dust mass distribution is that a large proportion (78 \%) of the 9 radio galaxies
with the highest dust masses ($M_d > 2\times10^7$~M$_{\odot}$) -- comparable with those of ULIRGs -- belong to the rare class of radio galaxies that
show evidence for prodigious star formation activity \citep{tadhunter11,dicken12}. Indeed, the two highest dust mass radio galaxies -- PKS2314+05 and PKS2135-20 -- are classified as ULIRGs based on their far-IR luminosities.
\newpage\noindent
\section{Discussion and conclusions}

The dust masses of the radio source host galaxies, and the total cool ISM masses that they imply, provide 
important clues to the nature of the events that triggered the powerful AGN and jet activity in these systems,  complementing existing studies of the 2Jy sample host galaxy morphologies \citep{ramos11,ramos12} and levels of star formation activity \citep{dicken12}.

The radio galaxies with the highest dust masses have cool ISM contents, morphologies and star formation properties that are
similar to those of ULIRGs in the local Universe. In such cases it is likely that the AGN have been triggered in major gas-rich mergers in which a substantial fraction of the stellar mass is being built up via merger-induced star formation and in
which the black holes are undergoing a major growth phase. However, such starburst radio galaxies are rare in the local radio
galaxy population \citep{tadhunter11,dicken12}. More typical radio galaxies have much lower cool ISM masses, and most of them lack evidence for recent star formation activity \citep{dicken12}, even if they do show morphological signs of recent galaxy interactions \citep{ramos11,ramos12}. In these more typical cases it is likely that we are witnessing the late-time re-triggering of quasar activity via mergers between 
mature giant elliptical galaxies and companion galaxies that have gas masses between those of the
LMC and M33 in the Local Group. As we showed in the introduction, such mergers are capable of delivering sufficiently
massive reservoirs of cool gas to the radio source hosts to fuel quasar-like activity on the requisite timescales.

Finally, we note that, while most of the nearby elliptical galaxies that lack powerful AGN activity have low dust masses (see Figure 2), a small but significant fraction of such objects have much larger dust masses. An interesting example in the local Universe is Centaurus A, which is associated with a relatively low-power FRI radio source and currently lacks a powerful quasar AGN, but nonetheless contains a substantial reservoir of cool ISM in its well-known dust lane \citep[$M_d = 1.6\times10^7$~M$_{\odot}$; $M_g = 2.7\times10^9$~M$_{\odot}$:][]{parkin12}. This raises the question of why Centaurus A, and gas-rich but quiescent elliptical galaxies in general, do not always appear as quasars. Most likely, the presence of a substantial gas reservoir is a necessary but not sufficient condition for triggering luminous quasar activity; other factors, such as
the detailed distribution and kinematics of the cool ISM are also likely to be important. In the particular case of Centaurus A it is plausible that we are observing the system in a late post-merger phase in which the ISM in the dust lane has settled into a dynamicallly stable configuration \citep[see also][]{tadhunter11}. In this case, the radial gas flows into the nuclear regions are insufficient to sustain luminous quasar activity. In contrast, objects with quasar nuclei may represent an earlier phase in the merger when the gas is still settling into a stable configuration, and the rate of radial flow of the gas into the nucleus is much higher. 
Clearly, high resolution CO observations that measure both the distribution and kinematics of the cool ISM in nearby radio galaxies with quasar-like nuclei will be crucial for making further progress in this field. 
Such observations will also provide an important check on the cool ISM masses 
that we have estimated from
the thermal dust emission.

\section*{Acknowledgments} We thank everyone involved with the Herschel Observatory and the PACS and SPIRE instruments. CT acknowledges support
from STFC consolidated grant ST/J001589/1. CRA is supported by a Marie Curie Intra European Fellowship within the 7th European Community Framework Programma (PIEF-GA-2012-327934). 
RM gratefully acknowledges support from the ERC under the European Union's Seventh Framework Programme (FP/2007-2013)/ERC Advanced Grant RADIOLIFE-320745.

\end{document}